\documentclass[prr,aps,showpacs,twocolumn,superscriptaddress,9pt]{revtex4-2}
\usepackage{hyperref} 
\usepackage{amsmath}
\usepackage{mathtools}
\usepackage{xcolor}
\usepackage{amsfonts}
\usepackage{amssymb}
\usepackage[percent]{overpic}
\usepackage{subcaption}
\usepackage{graphicx}
\usepackage{dsfont}
\usepackage{physics}
\usepackage[font=small,justification=raggedright]{caption}
\usepackage[version=4]{mhchem}
\usepackage[mathscr]{euscript}

\usepackage[figurename=FIG.]{caption}

\usepackage{ulem}

\begin{abstract}
Experimental platforms based on ultracold atomic gases have significantly advanced the quantum simulation of complex systems, yet the exploration of phenomena driven by long-range interactions remains a formidable challenge. 
Currently available methods utilizing dipolar quantum gases or multi-mode cavities allow to implement long-range interactions with a $1/r^3$ character or with a spatial profile fixed by the mode-structure of the vacuum electromagnetic field surrounding the atoms, respectively.
Here we propose an experimental scheme employing laser-painted cavity-mediated interactions, which enables the realization of atom-atom interactions that are fully tunable in range, shape, and sign. Our approach combines the versatility of cavity quantum electrodynamics with the precision of laser manipulation, thus providing a highly flexible platform for simulating and understanding long-range interactions in quantum many-body systems. Our analytical predictions are supported by numerical simulations describing the full dynamics of atoms, laser, and cavity. The latter demonstrate that there is a wide and experimentally accessible parameter regime where our protocol optimally works. The methodology not only paves the way for exploring new territories in quantum simulation but also enhances the understanding of fundamental physics, potentially leading to the discovery of novel quantum states and phases.
\end{abstract}

\begin{document}
\title{Laser-painted cavity-mediated interactions in a quantum gas}
\author{Mariano \surname{Bonifacio}}
\affiliation{\small Max-Planck-Institut f\"ur Physik komplexer Systeme, 01187 Dresden, Germany}

\author{Francesco \surname{Piazza}}
\email[]{piazza@pks.mpg.de}
\affiliation{\small Max-Planck-Institut f\"ur Physik komplexer Systeme, 01187 Dresden, Germany}
\affiliation{\small Theoretical Physics III, Center for Electronic Correlations and Magnetism,
Institute of Physics, University of Augsburg, 86135 Augsburg, Germany}

\author{Tobias \surname{Donner}}
\affiliation{\small Institute for Quantum Electronics, Eidgen\"ossische Technische
Hochschule Z\"urich, Otto-Stern-Weg 1, CH-8093 Zurich, Switzerland}

\maketitle

Quantum simulators hold the potential to unravel intricate quantum states and dynamics of many-body systems that are challenging to analyze using traditional methods and impractical for classical numerical simulations~\cite{Feynman1982Simulating}. Specifically ultracold atoms have been established as a powerful experimental platform for simulating models ranging from condensed matter~\cite{Bloch2012Quantum,Gross2017Quantum,Schafer2020Tools} over high-energy physics~\cite{Zohar2016Quantum} to cosmology~\cite{Viermann2022Quantum}. This success is based on the almost perfect control over internal and external degrees of freedom of the particles, paired with their fully tunable local interactions provided by elastic atomic collisions.

A large number of phenomena in nature however arises due to long-range interactions between its constituents or due to competing long- and short-range interactions, and are thus out of reach for the traditional tool set of simulations using quantum gases~\cite{Defenu2021Long}. For example the formation of crystalline or glassy phases \cite{dalmonte2010LLstaircase,pupillo2010Rydberg,Cinti2014Defect,Mendoza2015Stripe,Rossotti2017Soft,Angelone2016Superglass,Cinti2019Thermal,Pupillo2020Quasi}, the existence of exotic many-body phenomena such as topological or frustrated magnetism~\cite{Micheli2005Toolbox,VanBijnen2015Quantum,Samajdar2021Quantum,Yao2013Realizing}, supersolidity \cite{Scarola2005Quantum,Leonard2017Supersolid,Tanzi2019Observation,Bottcher2019Transient,Chomaz2019Long} and roton excitations\cite{Santos2003Roton,Mottl2012Roton}, or quantum droplets~\cite{Chomaz2016Quantum,Ferrier-Barbut2016Observation,Cabrera2018Quantum}
usually require the presence of long-range interactions.
Not only interactions with a simple power-law decay $1/r^\alpha$ as a function of distance $r$ between the particles are of interest, with for example $\alpha=1$ for Coulomb interactions and $\alpha=3$ for dipolar interactions, but specifically also more complex interactions such as the Ruderman-Kittel-Kasuya-Yosida (RKKY)-type interaction~\cite{Kasuya1956Theory,Yosida1957Magnetic,Ruderman1954Indirect,Arguello-Luengo2022Tuning} which oscillates in space and is a fundamental model in condensed matter physics also applied to describe giant magnetoresistance.

Recently, tunable spin-exchange interactions have been implemented in ion chains and in coupled thermal atomic ensembles~\cite{PRXQuantum.1.020303,Periwal2021Programmable}. However, approaches to engineer long-range interactions in quantum gases usually involve  atoms with large magnetic dipole moments~\cite{Chomaz2023Dipolar,Bottcher2021New}, polar molecules~\cite{Baranov2012Condensed}, and Rydberg atoms~\cite{Adams2020Rydberg}, which all lead to a $1/r^3$ scaling of the interactions due to the underlying dipole-dipole coupling. An alternative approach to engineer long-range interactions in quantum gases makes use of hybrid systems combining quantum gases with high-finesse optical cavities, where photons from an external drive effectively mediate interactions between the involved atoms~\cite{Mivehvar2021Cavity}. Microscopically, a photon from the drive field impinging on the atomic cloud is scattered into the cavity mode at a first atom and then back into the drive at a second atom, mediating the effective interactions. For a single-mode cavity, where the photon is delocalized over the entire cavity mode, this results in global-range interactions which have been extensively exploited to realize lattice supersolids~\cite{Baumann2010Dicke,Klinder2015Dynamical}, quantum optics spin models~\cite{Ferri2021Emerging}, or extended Hubbard-models~\cite{Klinder2015Observation,Landig2016Quantum}. Coupling simultaneously to two cavity modes allowed the realization of a supersolid breaking a continuous translational symmetry~\cite{Leonard2017Supersolid}, and extending the coupling to many degenerate modes enabled the realization of finite-range~\cite{Vaidya2017Tunable} and sign-changing interactions~\cite{Guo2019Sign}, and the simulation of phonons in a lattice model~\cite{Guo2021Optical}. 
However, while the cavity-mediated interaction range becomes indeed tunable to some degree, their shape is still dictated by the involved cavity mode patterns and thus fixed by a given choice of cavity geometry.

In this article, we introduce a novel experimental scheme based on laser-painted cavity-mediated interactions, enabling the realization of atom-atom interactions fully tunable in range, shape, and sign. We detail the basic  setup and establish a comprehensive framework for the effective atom-atom interaction, encompassing both cavity-mediated long-range and collisional short-range interactions. Through numerical simulations in 1D, we illustrate the practicality, resilience, and boundaries of our proposed methodology. Our approach includes an integrated tomographic feature for the in-situ analysis of the atomic density profile through the cavity's output. The proposed scheme offers a gateway to explore the diverse realms of long-range interaction physics, not only making quantum simulations of purely theoretical models feasible but also opening doors to uncharted phenomenological exploration.

\section*{Results}
\subsection*{Experimental scheme}
\begin{figure}
\includegraphics[width=\columnwidth]{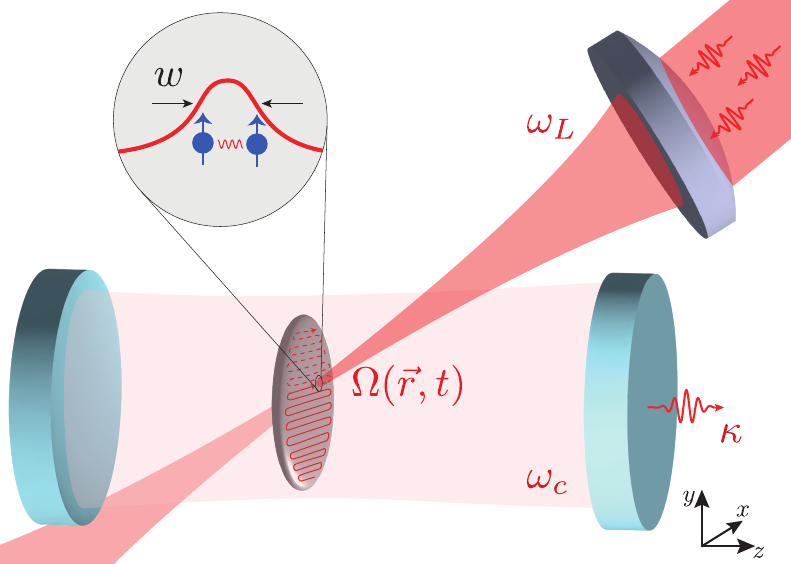}
\caption{Schematic of the cavity-QED experimental setup for the implementation of finite-range light-mediated interactions between atoms. A Bose-Einstein condensate (grey, not to scale) is strongly coupled to an optical cavity mode (light red mode confined by blue mirrors) with resonance frequency $\omega_c$, and tightly confined in a plane perpendicular to the axis of the cavity. The atomic cloud is driven by a focused laser field with mode function $\Omega(\vec{r},t)$ and frequency $\omega_L$ that scans across the atomic cloud periodically following a path of horizontal lines. Photons leaking from the cavity mode at rate $\kappa$ provide real-time information on the atomic system. The inset shows the microscopic process of light-mediated interactions: Only atoms within the spot illuminated by the transverse laser field interact with each other via the exchange of photons. The interaction potential is determined by the shape of the laser spot. Time-averaging of the process results in a effective atomic interaction in the entire cloud.}
\label{fig:Fig1}
\end{figure}

The experimental concept is visualized in Figure~\ref{fig:Fig1}. A Bose-Einstein condensate (BEC) is dispersively coupled with single-atom vacuum Rabi rate $g_0$ to a single optical cavity mode with resonance frequency $\omega_c$, wavelength $\lambda_c$, and field decay rate $\kappa$. The oblate BEC with lateral size $L$ is confined in a plane orthogonal to the cavity axis such that it resides at an anti-node of the cavity mode and the coupling to the cavity is approximately constant across the entire cloud. Also schemes using a different geometry are feasible as we demonstrate with our numerical simulations. The BEC is illuminated by a focused laser field with waist $w \ll L$ and  local Rabi frequency $\Omega(\vec{r},t)$. This laser spot is swept as a function of time across the BEC following a path $\Vec{r}_0(t)$ along horizontal lines separated by a distance $l$.

Concentrating for the moment only on the region illuminated by the laser beam at a certain moment in time, effective interactions between two atoms are generated by a cavity-mediated dipole-dipole coupling \cite{Mivehvar2021Cavity}. Since the frequency $\omega_L$ of the laser is chosen far red detuned by $\Delta_a = \omega_a - \omega_L \gg \Gamma$ with respect to the atomic resonance frequency $\omega_a$, where $\Gamma$ is the decay rate of the atomic transition, the atoms are not electronically excited, but an oscillating electric dipole moment is induced in each atom. A free-space coupling between two induced dipoles had been suggested as a mechanism to generate long-range interactions~\cite{ODell2003Rotons,Giovanazzi2002Density}. These direct dipole-dipole interactions decay with $1/r^3$, and in addition exhibit the same  $(\Gamma/\Delta_a)^2$ scaling as spontaneous scattering~\cite{Grimm1999Optical}. As a result, interesting quantum phases induced by this interaction are endangered by heating \cite{Ostermann2016Spontaneous,Dimitrova2017Observation}.
The presence of a strongly coupled cavity mode however boosts the interaction between the induced dipoles by the finesse of the cavity \cite{PhysRevLett.84.4068}:
The field radiated by an atomic dipole will predominantly drive the cavity mode if the detuning $\Delta_c=\omega_c - \omega_L$ is sufficiently small. The cavity field then will interfere with the driving laser field at the position of the other atomic dipole, mediating an interaction between the two. Only atoms simultaneously coupled to the cavity and illuminated by the laser spot will participate in this coupling, since the cavity field alone is too small to significantly influence the dipole moment of an atom without amplification by the interference term. 

This process induces for $\Delta_a,\Delta_c<0$ attractive ``local" interactions between atoms at the position of the laser spot, and repulsive ones for $\Delta_c>0$. The range and shape of these interactions is set by the field distribution $\Omega(\vec{r},t)$ of the laser mode, which can be freely engineered in the experiment using beam shaping techniques. Even far distant atoms can be made to interact if two regions of the BEC are simultaneously illuminated. 

In order to generate a translational invariant interaction between the atoms in the entire BEC, the focused beam is rapidly scanned across the cloud at rate $\omega_\mathrm{scan}$. Analogous to the creation of time-averaged optical potentials~\cite{Schnelle2008Versatile}, this process generates a time-averaged interaction between the atoms. Two time scales are important when considering this scheme: Atomic density patterns emerge and decay on a time scale given by the recoil frequency $\omega_R=2\pi^2/M\lambda_c^2$, such that the beam has to be scanned across the cloud much faster than this time scale, with $M$ being the atomic mass. On the other hand, the cavity decay rate $\kappa$ determines if the field scattered into the cavity by the atoms illuminated at one position has decayed before the spot has moved to the next position. This is required such that the interaction is only induced where the scanning laser spot resides. If the beam scans across the cloud before the field decays, the system would behave the same as a globally illuminated cloud, i.e. giving rise to infinite-range interactions. Assuming typical values of $\omega_R\approx$ kHz and $\kappa \approx$ MHz \cite{Mivehvar2021Cavity}, leaves a broad frequency window for the scanning frequency $\omega_\mathrm{scan}$.

\subsection*{Effective atom-atom interaction}
\label{sec:derivation_general}
 We consider the situation sketched in Figure \ref{fig:Fig1} with $N$ atoms coupled to a single mode cavity and subject to a scanning laser field that is far red detuned from atomic resonance, $\Delta_a < 0$. With $\Delta_a$ being the largest energy scale, the excited state of the atoms can be adiabatically eliminated and the dynamics of the system is well described in mean-field approximation by the Gross-Pitaevskii equation for the condensate wavefunction $\psi(\vec{r},t)$, normalized to $N$ \cite{Mivehvar2021Cavity,piazza2013bose,lang2017collective}
\begin{equation}
\begin{split}
i\frac{\partial  {\psi}(\vec{r},t)}{\partial  t}  =\Big[-\frac{\hbar \nabla^{2}}{2 M}+\frac{g_{\text{aa}}}{\hbar}|{\psi}(\vec{r},t)|^2+\frac{\Omega^2(\vec{r},t)}{\Delta_a} +&\\ +\frac{g_0^2}{\Delta_a} |\alpha(t)|^2+\frac{g_0\Omega(\vec{r},t) }{\Delta_a}(\alpha(t)+\alpha^*(t)) \Big]  {\psi}(\vec{r},t)&
\end{split}
\label{eq:EoM_psi_1}
\end{equation}
which is coupled to the equation for the evolution of the coherent field amplitude $\alpha(t)$ of the cavity
\begin{equation}
\begin{split}
i\frac{\partial  \alpha(t)}{\partial  t} =&(-\Delta_c-i\kappa) \alpha(t)  +\\&+\frac{1}{\Delta_a} \int g_0[g_0 \alpha(t)  + \Omega(\vec{r},t) ] |{\psi}(\vec{r},t)|^2 d \vec{r}\,,
\end{split}
\label{eq:EoM_a}
\end{equation}
where $M$ is the atomic mass, $g_{\text{aa}}$ is the effective contact interaction strength between atoms associated with the s-wave scattering and ${\Omega(\vec{r},t)=\Tilde{\Omega}(\vec{r}-\vec{r}_0(t))}$ is the mode function of the laser beam.

We assume the characteristic time scale $1/\kappa$ of the cavity field to be fast with respect to the atomic motion and adiabatically eliminate the cavity field (see Methods). In addition, we consider a regime of parameters in which the movement of the laser is faster than the atoms' dynamics, this is when the angular frequency of the laser sweep is large compared to both the recoil frequency and the contact interaction ${\omega_\mathrm{scan} \gg \omega_R,\hbar^{-1}g_\text{aa}|\psi|^2}$. In this situation one can take the average over one period of the laser displacement $T=2\pi/\omega_\mathrm{scan}$, which results in the equation of motion
\begin{equation}
\begin{split}
i\frac{\partial  {\psi}(\vec{r},t)}{\partial  t}  =\Big[-\frac{\hbar \nabla^{2}}{2 M}+\frac{g_{\text{aa}}}{\hbar}|{\psi}(\vec{r},t)|^2+\frac{\overline{\Omega^2(\vec{r},t)}}{\Delta_a}+&\\+\int V(\vec{r},\vec{r}\,')|{\psi}(\vec{r}\,',t)|^2 d\vec{r}\,' \Big]  {\psi}(\vec{r},t)&
\end{split}
\label{eq:EoM_psi_3}
\end{equation}
with effective interaction
\begin{equation}
V(\vec{r},\vec{r}\,')=\frac{2\Delta_c g_0^2}{\Delta_a^2 (\kappa^2+\Delta_c^2) } \overline{\Omega(\vec{r},t)\Omega(\vec{r}\,',t)}\,,
\label{eq:V_Interaction}
\end{equation}
where
\begin{equation}
    \overline{\Omega(\vec{r},t)\Omega(\vec{r}\,',t)}=\frac{1}{T}\int_0^T dt \Tilde{\Omega}(\vec{r}-\vec{r}_0(t))\Tilde{\Omega}(\vec{r}\,'-\vec{r}_0(t))\,.
    \label{eq:average_1}
\end{equation}

The term $\frac{\overline{\Omega^2(\vec{r},t)}}{\Delta_a}$ in Equation (\ref{eq:EoM_psi_3}) is a constant position-independent potential for the atoms and does not contribute to the dynamics. For simplicity, we assume a centrosymmetric laser spot, $\Tilde{\Omega}(-(x,y))=\Tilde{\Omega}((x,y))$, for which the above expression takes the form of a convolution ${\frac{1}{L^2}(\Tilde{\Omega}\ast\Tilde{\Omega})(\vec{r}-\vec{r}\,')}$, see Methods. As a result, the atom-atom interaction in momentum space is determined by the square of the Fourier transform of the shape of the laser spot,
\begin{equation}
    \hat{V}(\vec{k})=\frac{2\Delta_c g_0^2}{\Delta_a^2 (\kappa^2+\Delta_c^2) } \frac{1}{L^2} \hat{\Tilde{\Omega}}(\vec{k})^2\,.
    \label{eq:V_Interaction_k_space}
\end{equation}
With the current experimental state of the art, this interaction potential can be essentially designed at will: Rapid scanning of a laser beam is routinely achieved using galvo-mirrors, polygon scanners, or acousto-optical deflectors, while arbitrary and even time dependent beam shapes can be produced using digital mirror devices (DMDs). DMDs are specifically suited if placed in the Fourier plane of the optics projecting the laser spot onto the atoms, such that the interaction potential, Equation (\ref{eq:V_Interaction_k_space}), can be implemented straightforward in momentum space.

Thanks again to the hierarchy of timescales $\kappa\gg \omega_\mathrm{scan} \gg \omega_R,Ng_{\text{aa}}/(\hbar L^2)$, the light field leaking from the cavity allows to efficiently probe the  atomic density profile in real-time. Indeed, since the laser field $\Omega$ changes much faster than the atomic wavefunction $\psi$, the former performs a full scan over the cloud before the latter has evolved. This tomographic signal is directly accessible as a time-modulation of the complex-valued cavity field.

\subsection*{Example in 1 dimension}

\begin{figure*}[tb]
\begin{minipage}[b]{0.46\linewidth}
\centering
\includegraphics[width=0.75\textwidth]{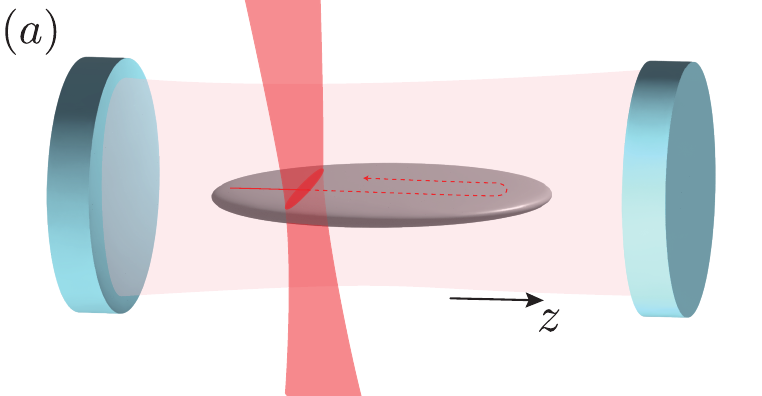} \\
\includegraphics[width=\textwidth]{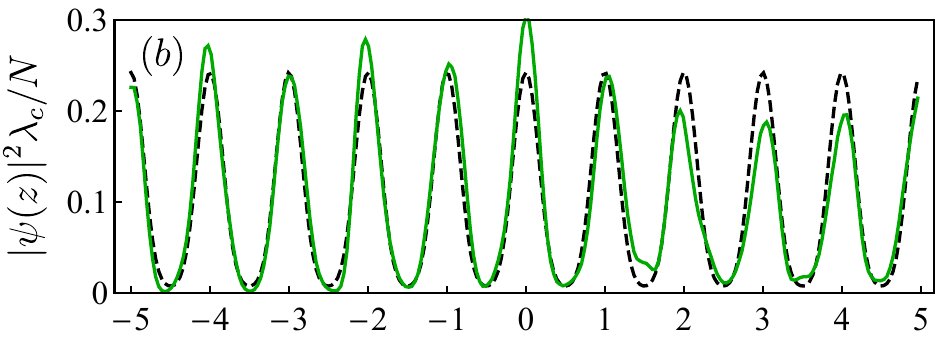} \\
\includegraphics[width=\textwidth]{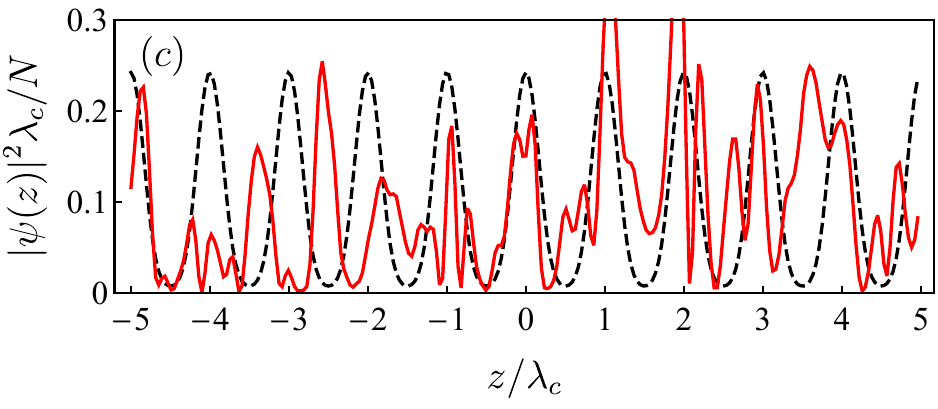}

\end{minipage}
\hspace{0.5cm}
\begin{minipage}[b]{0.44\linewidth}
\centering
\includegraphics[width=\textwidth]{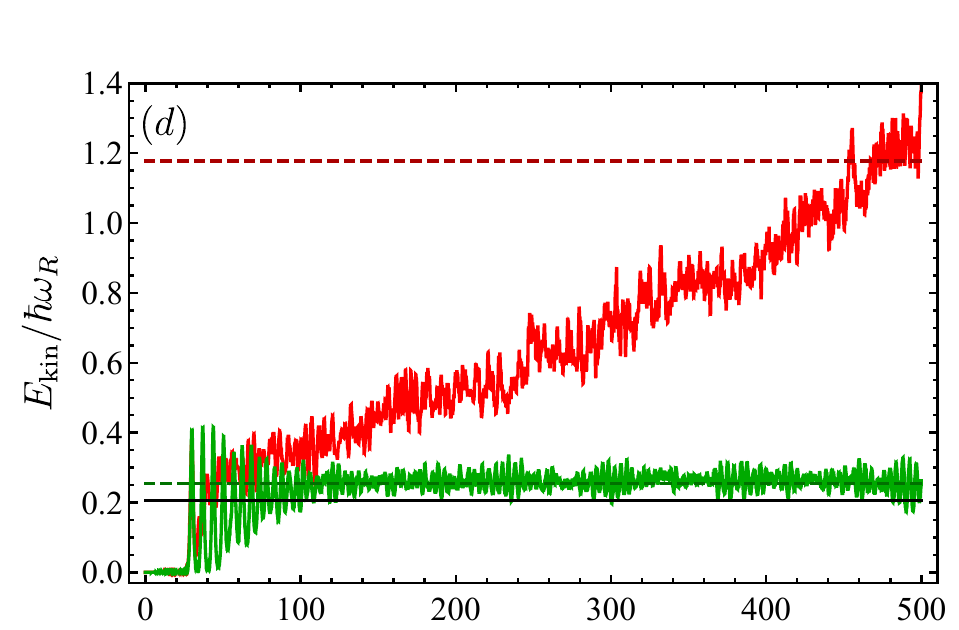} \\
\includegraphics[width=\textwidth]{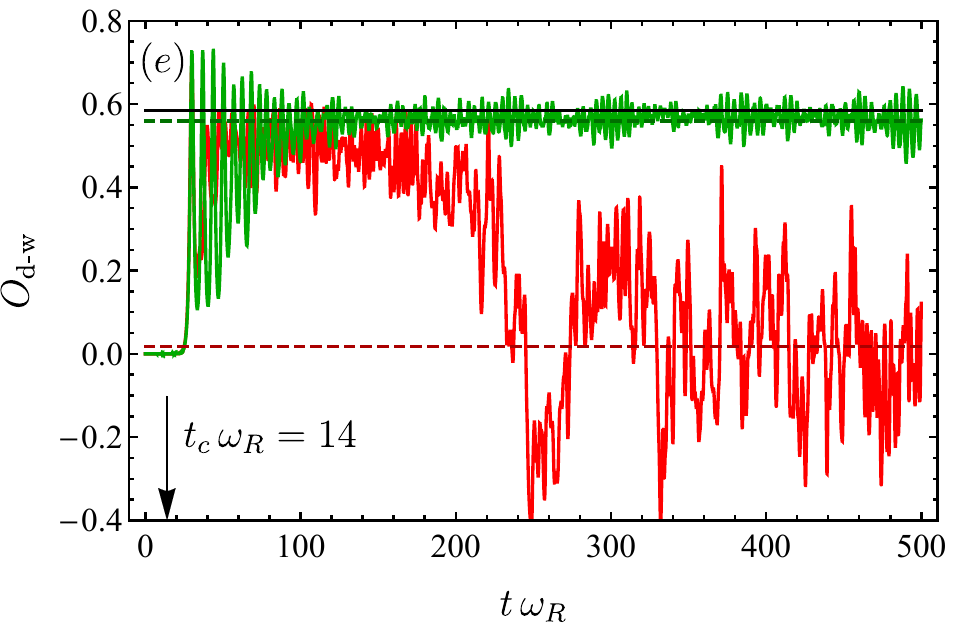}
\end{minipage}
\caption{(a) One dimensional realization of the protocol. An elongated cloud is coupled to a standing wave cavity and illuminated by a narrow laser beam which periodically scans along the $z$-direction across the cloud. (b-c) Comparison between the atomic density at time $t=500\omega_R^{-1}$ obtained by the real-time dynamics for scanning frequency of the laser $\omega_\mathrm{scan}=62.8\omega_R$ (green) and $\omega_\mathrm{scan}=628\omega_R$ (red) and the ground state obtained by the imaginary-time evolution with the effective interaction (dashed black) for parameters $L=10\lambda_c$, $w=1\lambda_c$, $N=10^5$, $g_{\text{aa}}=10^{-4}\hbar\omega_R\lambda_c$, $\tau=10\omega_R^{-1}$, and $V_0=-1.55\times10^{-4}\omega_R$. This value of the effective cavity-mediated interaction strength can be achieved with different combinations of the microscopic parameters, as long as the separation of time scales is fulfilled. In particular, we chose $g_0=600\omega_R$, $\kappa=250\omega_R$, $\Delta_c=-2500\omega_R$, $\Delta_a=-2.5\times10^7\omega_R$ and $\Omega_0=5.2\times10^4\omega_R$. Notice that rescaling $g_0$, $\Omega_0$ and $\sqrt{\Delta_a}$ by the same factor leaves the equations of motion invariant. (d-e) Kinetic energy and density-wave order parameter as a function of time (color solid lines) and the average of these quantities in the time interval $[450\omega_R^{-1},500\omega_R^{-1}]$ (dashed lines). The solid black line indicates the value obtained for the ground state by imaginary-time evolution. The time $t_c$ indicates the moment at which the critical intensity of the laser is reached for the transition from the homogeneous to the modulated phase.}\label{fig:Fig2}
\end{figure*}

In order to demonstrate the protocol and to test its regime of validity, we focus on one of the simplest possible situations, which further is experimentally realizable in a straight-forward way.  We consider a quasi-1D cigar-shaped cloud aligned with the axis of the cavity. The derivation provided in the previous section remains valid, but one needs to replace $g_0\rightarrow g(\vec{r})=g_0\cos(2\pi z/\lambda_c)$ to take into account the spatial modulation of the cavity mode. We employ a Gaussian laser profile $\Tilde{\Omega}(z)=\Omega_0 e^{-z^2/w^2}$, for which the effective interaction (\ref{eq:V_Interaction}) takes the form
\begin{equation}
\label{eq:effective_int_1D}
    V(z,z')=V_0\cos(2\pi z/\lambda_c)\cos(2\pi z'/\lambda_c)e^{-|z-z'|^2/2w^2}
\end{equation}
with 
\begin{equation*}
    V_0=\frac{\sqrt{2\pi}\Delta_c g_0^2\Omega_0^2}{\Delta_a^2 (\kappa^2+\Delta_c^2) }\frac{w}{L}\,.
\end{equation*}

For negative values of $\Delta_c$ this interaction is attractive and competes with the repulsive contact interaction and the kinetic energy. For small values of $V_0$, the ground state of the system thus remains homogeneous. However, by increasing $V_0$ it can undergo a phase transition to a phase with periodically modulated density. We compute the ground state resulting from the effective interaction \eqref{eq:effective_int_1D} by evolving Eq. (\ref{eq:EoM_psi_3}) in imaginary time. In order to verify that our laser-painting protocol does reproduce the physics of the effective atom-only model (\ref{eq:EoM_psi_3}), we compare the latter with the full dynamics obtained from the evolution of Eqs. (\ref{eq:EoM_psi_1}) and (\ref{eq:EoM_a}) in real time. As initial conditions, we choose a homogeneous atomic density and zero cavity field. The laser intensity is ramped up according to $\Tilde{\Omega}(z,t)=\Omega_0\tanh(t/\tau)e^{-z^2/w^2}$, and periodic boundary conditions are employed.

We analyze the transition between the homogeneous and the modulated phase with help of the kinetic energy per atom
\begin{equation}
    E_\text{kin}=\frac{1}{N}\frac{\hbar^2}{2M}\int_{-L/2}^{L/2}\,dz |\partial_z\psi(z)|^2
\end{equation}
and the density-wave order parameter capturing the overlap of the atomic density with a perfect $\lambda_c$-periodic modulation favoured by the long-range interaction,
\begin{equation}
\label{eq:dw_op}
    O_{\text{d-w}}=\frac{1}{N}\int_{-L/2}^{L/2}\,dz \cos(2\pi z/\lambda_c) |\psi(z)|^2.
\end{equation}

Figure \ref{fig:Fig2} characterizes the protocol, comparing two different regimes for the scanning rate $\omega_\mathrm{scan}$ of the laser spot. For intermediate rates (green lines), $\omega_\mathrm{scan}=62.8\omega_R$, the protocol works as intended. Initially, the system is in the homogeneous state, but some time after the laser strength has been ramped up across the critical value, the atomic system transitions to the density-wave phase, showing a modulation in agreement with one found in the ground state of the effective model (\ref{eq:EoM_psi_3}) (black-dashed lines in panel (b)). Residual deviations are due to the finite-time ramp and are present also within the effective model by comparing the long-time state with the ground state. This is demonstrated in panels (d) and (e), showing the time evolution of kinetic energy and density-wave order parameter as green solid lines. Their long-time average has a small deviation from the ground-state value (solid black line), but coincides with the long-time average computed using the effective model (\ref{eq:EoM_psi_3}) (green dashed line).  

For large scanning rates instead, $\omega_\mathrm{scan}=628\omega_R$, the cavity is unable to follow adiabatically the dynamics of the laser (red lines). The combined dynamics of cavity and laser now has beatings at a frequency much slower than the respective characteristic timescales, which can in turn efficiently excite the atoms. These beatings show up in the oscillations of the laser-cavity-interference potential felt by the atoms (last term in Equation \eqref{eq:EoM_psi_1}). This leads to the heating manifested as monotonic growth of the kinetic energy (red line in Fig. \ref{fig:Fig2}(d)). Correspondingly, the atomic density gets modulated at multiple wavelengths and thus the density-wave order parameter \eqref{eq:dw_op} decreases and oscillates around a value that vanishes in the thermodynamic limit, as more and more wavelengths (momenta) get excited.

\begin{figure}
\begin{center}
\includegraphics[width=1\columnwidth]{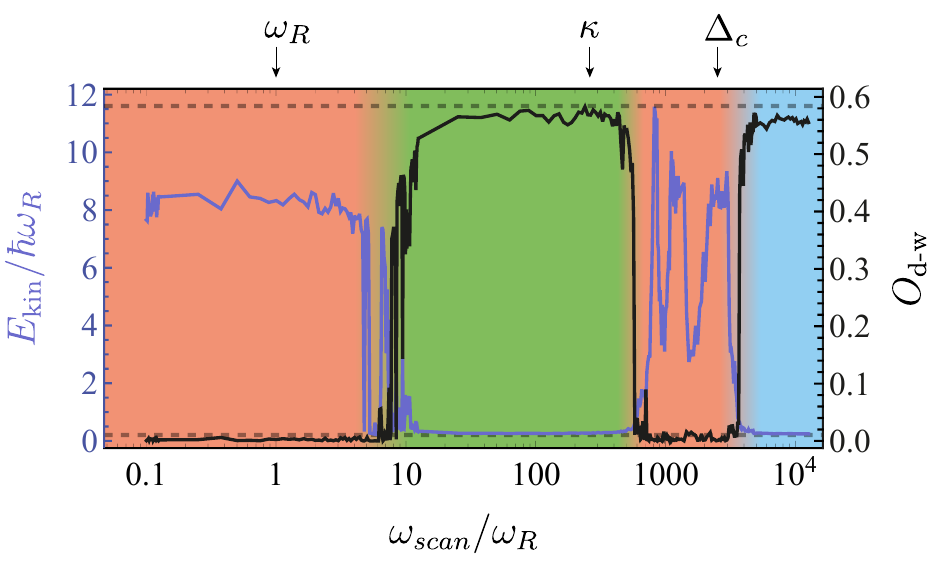}
\end{center}
\caption{Kinetic energy per particle (blue) and order parameter (black) averaged in the time interval $[450\omega_R^{-1},500\omega_R^{-1}]$ for different scan rates of the laser, considering the same parameters as in Fig. \ref{fig:Fig2}. The dashed lines indicate the value obtained for the ground state of the static effective model \eqref{eq:effective_int_1D}. The scan-rate parameter range in which our laser-painted interaction protocol works is indicated by the green background. Scan rates which lead to heating (increased kinetic energy) are indicated by the orange background, while the parameter range in which the interaction becomes of global range is indicated by the blue background.}
	\label{fig:Fig3}
\end{figure} 

In order to provide the full characterization of the protocol with respect to the robustness of the hierarchy of timescales, we show in Figure \ref{fig:Fig3} the kinetic energy and order parameter averaged over the time interval $[450\omega_R^{-1},500\omega_R^{-1}]$ as a function of the laser scanning rate $\omega_\mathrm{scan}$. 
For slow scan rates ($\omega_\mathrm{scan} \ll \omega_R$), we observe the formation of a single peak in the density produced by the negative potential contribution $\Omega(\vec{r},t)^2/\Delta_a$ in equation~\eqref{eq:EoM_psi_1}, which decays once the laser moves on. Since the system is perturbed in all its length during the ramping of the laser intensity, a strong increment of the kinetic energy that persists in time is induced. For very fast scan rates ($\omega_\mathrm{scan} \gg \Delta_c$), however, the laser acts like it is spatially constant, resulting in a modulation of the atomic density that cannot be distinguished from global cavity-mediated interactions. For scan rates comparable to one of the natural time scales of the system ($\omega_R, \kappa$,$\Delta_c$), energy is efficiently transferred to the atoms, either due to Floquet heating induced by the local laser beam moving periodically at a rate $\omega_{\rm scan
}\sim \omega_R$, or due to the excitation of the cavity by the laser whenever $\omega_{\rm scan}$ is comparable to a cavity timescale. As discussed above (see Fig.\ref{fig:Fig2}), in the latter case heating is caused by beating oscillations in the cavity-laser interference potential. A plateau of very small kinetic energy and large order parameter, approaching the respective ground-state values, appears in a broad range of intermediate scan rates, demonstrating the robustness of our protocol. For the parameters employed in the figure, this plateau lies roughly between $15\omega_R$ and $500\omega_R$. We note also that, when the cavity loss rate is by far the largest scale, cavity cooling/heating is inefficient \cite{piazza2014quantum}. 

\subsection*{Cavity-based tomography of domains}

The light leaking from the cavity mirrors can be used to obtain tomographic real-time information about the density of the atoms, which we demonstrate   using the example from the previous section. Applying the cavity field elimination (see Methods) to the 1D situation, we obtain
\begin{equation}
    \alpha(t)=\frac{ \int dz g_0\cos(2\pi z/\lambda_c)\Omega(z,t) |{\psi}(z)|^2 }{\Delta_a(\Delta_c+i\kappa)}\,,
    \label{eq:cavity_tomography}
\end{equation}
where the atomic density is considered static as it evolves with the slowest time scale. 
In the density-wave phase, the atoms spontaneously break a  $\mathbb{Z}_2$ symmetry \cite{Mivehvar2021Cavity} with density maxima or minima at integer values of $z/\lambda_c$. If the range of the interaction is sufficiently smaller than the size of the system ($w \ll L$), domains with different spontaneous breaking of the symmetry can be formed as the critical point is crossed within a finite time. According to the Kibble-Zurek mechanism \cite{kibble1976topology,zurek1985cosmological}, we thus expect in our case the formation of domains with different position of the density-modulation maxima when we ramp the laser strength across the homogeneous-to-density-wave critical point. This is in stark contrast to a situation with global-range interactions which does not allow for any domain formation.

\begin{figure}[htp]
\begin{subfigure}[ht!]{0.5\textwidth}
\centering
    \begin{overpic}[width=0.9\textwidth]{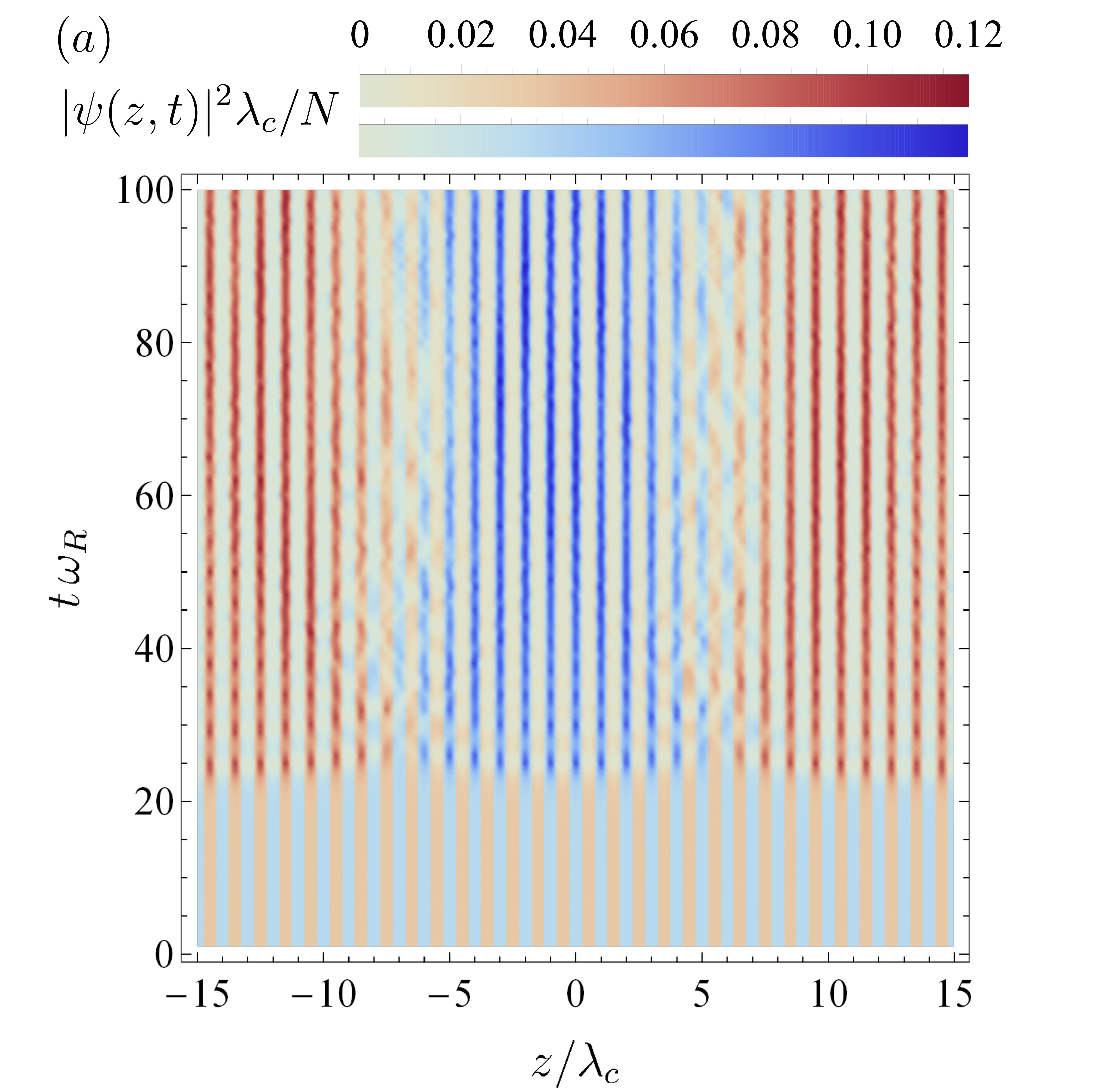}
    \end{overpic}
\end{subfigure}
\begin{subfigure}[ht!]{0.5\textwidth}
\centering
    \begin{overpic}[width=0.9\textwidth]{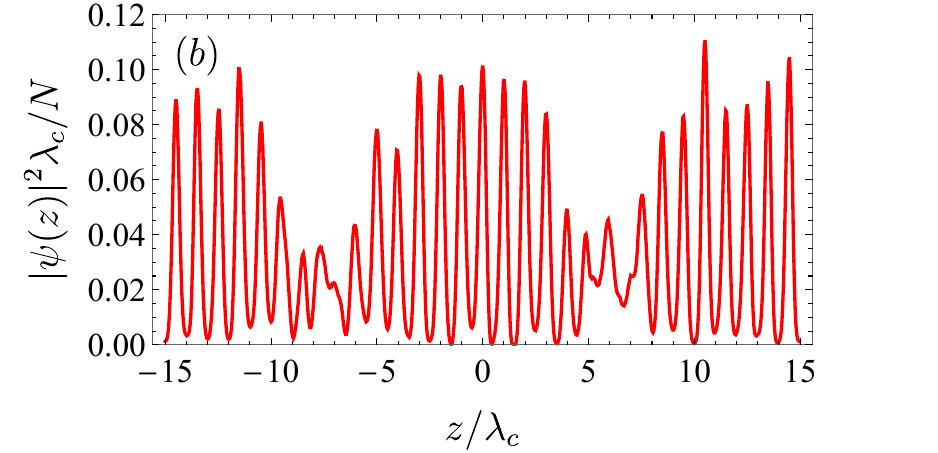}
    \end{overpic}
\end{subfigure}
\begin{subfigure}[ht!]{0.5\textwidth}
\centering
    \begin{overpic}[width=0.9\textwidth]{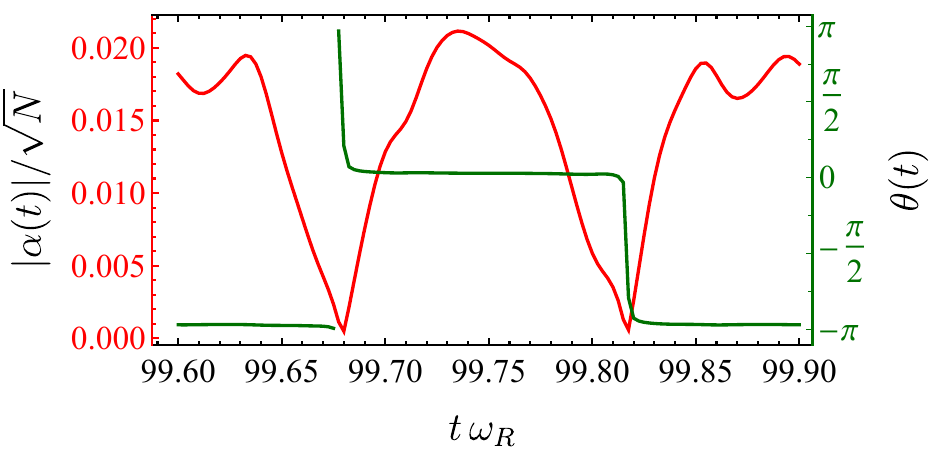}
    \put(18.5,42){(c)}
    \end{overpic}
\end{subfigure}

\caption{(a) Atomic density as a function of time starting from the homogeneous density for parameters $L=30\lambda_c$, $w=1\lambda_c$, $N=10^5$, $g_{\text{aa}}=3\times10^{-4}\hbar\omega_R\lambda_c$, $V_0=-4.9\times10^{-4}\omega_R$, $\tau=10\omega_R^{-1}$, and $\omega_\mathrm{scan}=20.9\omega_R$. The blue scale is used at integer positions and the red one at semi-integer positions. (b) Atomic density profile at time $t=100\omega_R^{-1}$. (c) Amplitude and phase of the cavity mode $\alpha$ as a function of time during one period of the laser sweep.}
	\label{fig:Fig4}
\end{figure}

The dynamics of domains can be characterized by analysing the cavity field $\alpha(t)$ during one period of the laser sweep. Figure \ref{fig:Fig4} shows results computed with the same parameters as in the previous section, but for a larger system, $L=30\lambda_c$, scanning frequency $\omega_\mathrm{scan}=20.9\omega_R$ and slightly different interaction strengths: $g_\text{aa}=3\times10^{-4}\hbar\omega_R\lambda_c$ 
and $V_0=-4.9\times10^{-4}\omega_R$.
As shown in panels (a) and (b), the atomic wave function develops two long-lived domains.

How the information about domains is stored in the cavity field dynamics can be understood analytically. By approximating the atomic-density modulation as $|\psi(z)|^2=(2N/L)\cos(\pi z/\lambda_c)^2$ or $|\psi(z)|^2=(2N/L)\sin(\pi z/\lambda_c)^2$ for the phase with maxima or minima at integer values of $z/\lambda_c$, respectively, we obtain  
\begin{equation}
    \alpha=\pm\frac{g_0\Omega_0\sqrt{\pi}wN}{\Delta_a(\Delta_c+i\kappa)2L}(1+\mathcal{O}(e^{-\pi^2 w^2/\lambda_c^2}))
\end{equation}
where $\pm$ corresponds to the 2 different cases. The time phase of the cavity field is thus shifted by $\pi$ as the laser moves from one domain to the other. 
This is confirmed in Fig. \ref{fig:Fig4} (c) where the amplitude takes the same value in both domains but the phase is shifted by $\pi$. Naturally, the amplitude vanishes when the laser is illuminating the region between domains. 
For this simple case of a Gaussian laser beam, the spatial resolution of the cavity tomography is set by the laser spot itself, i.e. by $w$. While cavity microscopes rely on overlapping multiple cavity modes for increasing the resolution \cite{kroeze2022highcooperativity}, our cavity tomograph relies on focusing the laser beam instead. In both devices, the resolution is at the same time the range of the interaction, and thus sets also the minimum size of the structures to be probed.

\section*{Discussion}
We theoretically demonstrated that the combination of a single-mode Fabry-Pérot cavity and a focused, scanning laser allow to implement effective interactions between atoms which are in principle arbitrarily tuneable in shape, strength, and sign.  Our protocol is well within reach of current experimental capabilities and also features a built-in tomographic probing tool based on the cavity output signal. It is important to stress at this point that the whole protocol works equally also for fermionic atoms \cite{zhang2021observation,helson2023density,mivehvar2017superradiant}.

As an experimentally realizable proof of principle, we characterized the protocol for the case of a Gaussian laser beam sweeping across a quasi-1D BEC trapped along the axis of a standing-wave cavity. This realizes a periodically sign-changing interaction, with a Gaussian envelope setting an interaction range given by the laser spot. We observed a transition between a homogeneous and a density-modulated BEC, the crossing of which generates different domains that can be probed in a space- and time-resolved fashion by the built-in tomographic tool.

Since the interaction potential can be freely painted pixel-by-pixel using digital beam shaping devices for the laser beam, the range of possibilities for future investigations is hard to delimit. Elusive ground states of matter that impose very specific requirements on the interaction potential can be stabilized and studied, including for instance strongly-correlated solitons and droplets induced by light \cite{fraser2019topological,karpov2022light}, quantum cluster quasi-crystals \cite{Pupillo2020Quasi,mivehvar2019emergent}, requiring very specific combinations of lengthscales to appear in the potential, or quantum paramagnets \cite{mivehvar2019cavity,chiocchetta2021cavity}, relying on strongly frustrated interactions in concomitance with specific symmetries. Our scheme thus allows the experimental implementation of so-far-purely-theoretical interaction potentials of paradigmatic nature, but might also lead to the discovery of yet-unknown phenomenology.

\section*{Methods}
\subsection*{Elimination of the cavity field}
If the characteristic time scale $1/\kappa$ of the cavity field is fast with respect to the atomic motion, the cavity field can be adiabatically eliminated which results in 
\begin{equation}
\alpha(t) \approx \frac{ \int g_0\Omega(\vec{r},t) |{\psi}(\vec{r},t)|^2 d \vec{r}}{\Delta_a(\Delta_c+i\kappa)}\,.
\label{eq:a_ss}
\end{equation}
Inserting this expression into Equation (\ref{eq:EoM_psi_1}) we find an effective description of the atomic subsystem
\begin{equation}
\begin{split}
&i\frac{\partial  {\psi}(\vec{r},t)}{\partial  t}  =\Big[-\frac{\hbar \nabla^{2}}{2 M}+\frac{g_{\text{aa}}}{\hbar}|{\psi}(\vec{r},t)|^2+\frac{\Omega^2(\vec{r},t)}{\Delta_a}+ \\&+\frac{2\Delta_c g_0\Omega(\vec{r},t)}{\Delta_a^2 (\kappa^2+\Delta_c^2) }\int g_0 \Omega(\vec{r}\,',t) |{\psi}(\vec{r}\,',t)|^2 d\vec{r}\,' \Big]  {\psi}(\vec{r},t)\,,
\end{split}
\label{eq:EoM_psi_2}
\end{equation}
where we used $|\Delta_a \Delta_c|\gg g_0^2 N$, and neglected $\frac{g_0^2}{\Delta_a}|\alpha|^2$ given the low occupation of the cavity mode. From here it is apparent that two atoms at positions $\vec{r}$ and $\vec{r}\,'$ only interact if they are simultaneously illuminated by the laser.

\subsection*{Averaging over the scan}
The laser has velocity $v$ and moves periodically through a path of horizontal lines separated by a distance $l$. In this way the period is $T=L^2/lv$ and the  average (\ref{eq:average_1}) can be parameterised as
\begin{equation}
    \frac{1}{T}\sum_{n=1}^{L/l}\int_{t_{n-1}}^{t_{n}} dt \Tilde{\Omega}(\vec{r}-\vec{r}_0^{\,n}(t))\Tilde{\Omega}(\vec{r}\,'-\vec{r}_0^{\,n}(t))\,,
    \label{eq:average_2}
\end{equation}
where the path for each line is given by ${\vec{r}_0^{\,n}(t)=(x_n(t),y_n)=(-\frac{L}{2}+v(t-t_{n-1}),-\frac{L}{2}+ nl)}$ and the times are $t_n=nL/v$. With the previous expression for the path, the integral in time can be written as an integral in space with the variable $x$. Furthermore, if the separation between lines is small compared to the characteristic size of the laser spot $w$,  the sum over $n$ can be turned into an integral in $y$ resulting in
\begin{equation}
    \frac{1}{L^2}\int_{-L/2}^{L/2} dy\int_{-L/2}^{L/2} dx\, \Tilde{\Omega}(\vec{r}-(x,y))\Tilde{\Omega}(\vec{r}\,'-(x,y))\,.
\end{equation}

Given that the laser has finite range, the integrand vanishes when the point $(x,y)$ is far away from $\vec{r}$ and $\vec{r}\,'$. Consequently, the integrals in space can be extended to $\pm\infty$ without changing the result. We can additionally make the change of coordinates $(x,y)\rightarrow\vec{r}\,'+(x,y)$ and obtain the expression 
\begin{equation}
    \frac{1}{L^2}\int_{-\infty}^{\infty} dy\int_{-\infty}^{\infty} dx\, \Tilde{\Omega}(\vec{r}-\vec{r}\,'-(x,y))\Tilde{\Omega}(-(x,y)).
\end{equation}

\section*{Acknowledgement}
We are thankful for insightful discussions with Alexander Baumgärtner, David Baur, Simon Hertlein, Gabriele Natale, and Justyna Stefaniak. T.D. acknowledges funding from the Swiss National Science Foundation SNF (Project No. IZBRZ2 186312), and funding from the Swiss State Secretariat for Education, Research and Innovation (SERI).


%

\end{document}